\documentclass[a4paper]{article}
\usepackage{amsmath}
\usepackage{graphicx} 
\usepackage{apalike}
\usepackage[a4paper, total={5.5in, 8in}]{geometry}
\usepackage[onehalfspacing]{setspace}

\title{Causal Emergence in Discrete \& Continuous Dynamical Systems}
\author{Thomas F. Varley $^{1,2}$}

\begin{document}
	\maketitle

	\textit{$^{1}$ Complex Networks \& Systems, School of Informatics, Computing \& Engineering, Indiana University, Bloomington IN, 47401, USA.}

	\textit{$^{2}$ Psychological \& Brain Sciences, Indiana University, Bloomington IN, 47401, USA.}
	
	\begin{flushleft}
		
		\textbf{Abstract}\newline 
		Emergence, the phenomena where a system's micro-scale dynamics facilitate the development of non-trivial, informative higher scales, has become a foundational concept in modern sciences, tying together fields as diverse as physics, biology, economics, and ecology. Despite it's apparent universality and the considerable interest, historically researchers have struggled to provide a rigorous, formal definition of emergence that is applicable across fields. Recent theoretical work using information theory and network science to formalize emergence in state-transition networks (causal emergence) has provided a promising way forward, however the relationship between this new framework and other well-studied system dynamics is unknown. In this study, we apply causal emergence analysis to two well-described dynamical systems: the 88 unique elementary cellular automata and the continuous Rossler system in periodic, critical, and chaotic regimes. We find that emergence, as well as its component elements (determinism, degeneracy, and effectiveness) vary dramatically in different dynamical regimes in sometimes unexpected ways. We conclude that the causal emergence framework provides a rich new area of research to explore both to theoreticians and natural scientists in many fields. 
	\end{flushleft}

	\section{Introduction}
		\label{sec:intro}
		The emergence of informative high scales in complex systems is a foundational concept in modern sciences. Fields from physics to life-sciences have explored how higher-order structures can emerge from lower-order dynamics, however complexity science as a whole still lacks a coherent, agreed-upon formalism that describes emergence, and more crucially, how much emergence a system is capable of displaying. Recently, Hoel et al. proposed an information-theoretic formalization of emergence based on the information encoded in a system's state-transition network \cite{hoel_quantifying_2013,hoel_can_2016}. Called causal emergence (CE), this framework provides a method to calculate the amount of information a system encodes in it's state-transition structure at the micro-scale and then, after renormalization, at the macro-scale. If a system has a more informative macro-scale than it's micro-scale, then it can be said to display "emergent structure." The the micro-scale is the most informative, then we say it displays "causal reduction."
		
		The majority of work using the causal emergence framework has been done using simple Boolean networks \cite{hoel_quantifying_2013,hoel_can_2016}, which can be solved through brute-force search. Klein \& Hoel \cite{klein_uncertainty_2019} applied CE analysis to a series of larger real and synthetic complex networks and found that the topology and structure of a given network can have a significant impact on it's capacity to encode information and emergence. However, a significant limitation of this work is that the majority of real networks explored are not readily understood as state-transition networks, instead often being network representations of interacting components (eg. gene networks, PGP web-of-trust), or routing networks (eg. airline networks, power-grids). We say this not to diminish the significance or validity of the work, but rather to draw attention to an outstanding question: \textit{given a discrete or continuous system with a well-defined state-transition network, what does CE analysis of the state-transition network tell us about the behaviour and dynamics of our system?} Without a paired system and it's state-transition graph, CE analysis provides limited intuitive understanding of what "emergence" might actually mean, beyond a kind of optimal community-detection schema.  
		
		To address this, we applied CE analysis (using the spectral-clustering algorithm proposed by Griebenow, Klein, \& Hoel \cite{griebenow_finding_2019} to two foundational classes of dynamical systems, each of which displays rich ranges of behaviour. The first, is the set of 88 unique discrete elementary cellular automata (ECA) \cite{wolfram_new_2002}. The ECA are an excellent starting point for this analysis as the different rules generate a wide range of behaviours that are easily visualized, including static, periodic, chaotic, and fractal-like regimes (described in more detail in Section  \ref{sec:eca}). In addition to this rich repertoire, for an ECA of a given size, all possible states the system can take on can be enumerated, and the transitions from state to state easily represented as a directed state-transition graph. Finally, as the ECA are deterministic in nature (a given state's immediate future can be predicted with total certainty), the information encoded in the causal structure is driven only by the degeneracy (determinism is constant). Previous work has found that high degeneracy is a key component of causal emergence \cite{hoel_can_2016}, and so the ECA allows us to explore that relationship without the added confound of varying determinism. 
		
		The second system we analysed was the Rossler system \cite{rossler_equation_1976}. A canonical model in chaos theory, the Rossler system is a set of coupled differential equations (described in detail in Section \ref{sec:rossler}) which can display periodic, chaotic, or critical behaviour depending on the values of it's parameters. By varying these values, we can explore the information-structure of the system as it passes through the phase transition from periodicity to chaos and back again. Unlike the ECA, however, the Rossler system is continuous and therefore doesn't naturally lend itself to this kind of discrete information-theoretic analysis. To construct a state-transition network for such a system requires a method of discretizing the attractor to create a finite number of states and the probabilities of transition from one to another. To do this, we used the method of constructing ordinal partition networks (OPNs), described in detail in \ref{sec:opns} \cite{small_complex_2013,mccullough_time_2015,myers_persistent_2019}. Briefly, an OPN represents the state-transition dynamics of a continuous time-series by representing discrete patterns of activity as nodes, and then counting the number of transitions from one state to another, which are stored as directed edge weights and can be normalized into probabilities for CE analysis. Unlike the ECA, which are perfectly deterministic, the OPN can have variable determinism and degeneracy depending on the dynamics of the system producing the source time-series. In many respects, the OPN is quite similar to the notion of the $\epsilon$-machine proposed by Crutchfield \cite{crutchfield_calculi_1994,crutchfield_between_2012}, which provides a provably optimal network representation of a continuous complex system. Once the OPN has been constructed, it can be used for the same analysis as the ECA state-transition network, although here determinism is variable, unlike in the ECA. By sweeping through the various dynamical regimes of the Rossler system and reconstructing a discrete attractor at every step, we can map system dynamics to a network topology for information-theoretic analysis. 
		
		The ECA and the Rossler OPN represent distinctly different kinds of dynamical system with many different behaviours and causal structures, but both are amenable for CE analysis. This provides an opportunity to relate well-understood dynamics such as chaos, periodicity, and complexity to the novel formalism of causal emergence and provide new insights into how dynamical systems encode information and why some may develop informative higher scales, while others do not. 
		
	\section{Materials \& Methods}
	\subsection{Causal Emergence Analysis}
	\label{sec:emergence}
	\subsubsection{Determinism, Degeneracy, \& Effectiveness}
	\label{subsec:det_deg}
	The causal structure of a state-transition graph is comprised of three fundamental elements: $determinism$, $degeneracy$ and $effectiveness$. Recall that for a weighted, directed state-transition graph $G=(V,E)$ where $|V| = N$, out-going edges are weighted by the probability of transitioning to a given future based on the current state and so satisfy a probability distribution. This allows us to define the determinism of the graph as a function of the average entropy of the distribution of out-going edges across all vertices:
	
	\[
	Det_2(G) = \frac{log_2(N) - \langle H(W^{out}_i) \rangle_{|i \in V}}{log_2(N)}
	\]
	
	The determinism gives an average measure of how reliably you can predict the future knowing the present. If every vertex in the graph has a single output with $p=1$, then the determinism of whole network is 1. Dividing by $log_2{N}$ introduces a normalizing fact which allows us to compare the determinism of networks with different numbers of vertices. 
	
	In contrast, the degeneracy is a function of the entropy of the "average" network distribution of out-going edges: 
	
	\[
	Deg_2(G) = \frac{log_2(N) - H( \langle W^{out}_i \rangle )_{|i \in V}}{log_2(N)}
	\]
	
	In this way, the degeneracy is the amount of information lost due to uncertainty about the past when multiple states could lead to the same present state. If every state had an equal probability of being preceded by any other state, the degeneracy is maximal at 1. If every state only had one possible past, then the network has 0 degeneracy. 
	
	The difference between the determinism and the degeneracy is the effectiveness: how much information does the network encode in it's causal structure. 
	
	\[
	Eff_2(G) = Det_2(G) - Deg_2(G)
	\]
	
	In a network with a very high effectiveness, determinism would be high (it is easy to predict future trajectories) with a low degeneracy (paths rarely overlap or feed into each-other). In a network with very low effectiveness, it is hard to both predict the future and reconstruct past. 
	
	\subsubsection{Renormalization \& Emergence}
	The formalism of effectiveness allows us to define an intuitive sense of what it means for a system to display "emergence": are there higher scales that have a greater effectiveness than the micro-scale? In the context of a defined state-transition graph, higher scales can be explored through "renormalization" where the micro-scale are coarse-grained by combining individual vertices into "macro-vertices". Renormalization is essentially a form of Louvain-like community detection, where the communities get collapsed into single vertices.  
	
	To implement the renormalization, we followed the algorithm detailed by Griebenow et al., \cite{griebenow_finding_2019}, which uses a modified form of spectral clustering to find causally similar vertices and assign them to a set. For the theoretical discussion of why this particular algorithm works, see the above-referenced paper. Briefly:
	\begin{enumerate}
		\item Calculate the eigendecomposition of the graph's adjacency matrix, resulting in a spectrum of typically complex eigenvectors ($V=\{v_i\}$) and their associated eigenvalues ($\Lambda=\{\lambda_i\}$). 
		\item Remove the kernel $K$ of the spectrum, where $K = \{v_i | \lambda_i = 0\}$.
		\item Rescale the remaining unit-length eigenvectors by the associated eigenvalue s.t. $V' = \{v_{i}^{'} | \lambda_iv_i\}$, creating a set of $N$-dimensional vectors.   
		\item We can associate the $i^{th}$ vertex in the network with a vector comprised of the $i^{th}$ elements of all the vectors in $V'$. This creates a set of $N$ vectors (one for each vertex in the network) that can be readily embedded in a space of dimension $N-|K|$.
		\item We then create a distance matrix for our $N$ vectors. If two vectors map to vertices that are within each-other's Markov blankets, we define the distance between them as the cosine distance between their associated vectors. If two vertices are not within each-other's Markov blankets, we define the distance to be $\infty$, as they should ever be clustered. 
		\item Finally, using our pre-computed distance matrix, we cluster the vertices using the OPTICS clustering algorithm \cite{ankerst_optics:_1999}. The output is a vector of assignments for each vertex. Nodes can either be mapped to a cluster (which get collapsed together into a macro-vertex), or treated as outliers, who remain independent in the renormalized network. 
	\end{enumerate}
	
	Once the macro-vertices have been assigned, the question remains: how do they fit into the rest of the network? We chose a relatively simple schema that erases information about the internal community structure while conserving as much information about the transition probabilities as possible. Briefly:
	
	\begin{itemize}
		\item The macro-vertex has one self-loop: it's weight is given by the average probability that a walker on any vertex within the community would transition to another vertex within the same community. 
		\item All in-coming edges incident on vertices within the macro-community terminate on the new macro-vertex, with the same probability. If multiple edges originate from the same source vertex and terminate on separate vertices in the macro-community, those edges are collapsed and their weights summed. 
		\item All out-going edges from vertices within the community originate from the new macro vertex. The weight of this edge is the average probability that a random walker on any vertex in the community would transition to the target vertex. 
	\end{itemize}
	
	The result of this procedure is that all information about the structure of the communities is lost after renormalization, however random walkers placed on both the micro- and macro-scale networks would follow similar paths. This method also ensures that, the weights of all the out-going edges for each vertex still sum to one. 
	
	\subsection{Elementary Cellular Automata}
		\label{sec:eca}
		Arguably the simplest cellular automata to explore are the 256 "Elementary Cellular Automata" (ECA), described in excruciating detail by Stephen Wolfram in his book "A New Kind of Science" \cite{wolfram_new_2002}. Each ECA consists of a one-dimensional array of cells, which can be in an on or off state. At every moment, each cell updates according to: it's current state, and the states of it's immediate left and right neighbours. The constraint that each cell's immediate future is totally specified by it's present environment makes these systems highly amenable to research. Considerable work has been done on the basic ECA (again, see Wolfram's 1,000 page tome on the subject) and a number of intriguing findings have emerged, including the discovery that at least one rule is a universal computer.
	
		For our purposes, the ECA are useful because they display a wide range of behaviours. Some, such as Rules 0 and 255 (the universal off and on rules respectively) produce trivial futures, regardless of their initial conditions, while others, like Rule 60 show highly non-trivial and complex behaviours. This allows us to ask: what kinds of behaviours are associated with causal emergence? At the outset, we were unsure: on one hand, the complex, fractal patterns that rules like Rule 60 produce are often used as examples of unexpected emergence, but on the other hand, each state in that case is highly individuated and it may not be easy to find a way to aggregate states in such variable systems. In contrast, simple patterns may be more compressible, but also less interesting. This analysis gives us an opportunity to test whether causal emergence in state-transition graphs corresponds with our own intuitions about what constitutes interesting or "non-trivial" emergence. Wolfram classified all of the ECA into four broad categories:\newline 
	
	\textbf{Classes of Elementary Cellular Automata}
	\begin{enumerate}
		\item Those ECA that converge to a uniform state. 
		\item Those ECA that converge to a stable, repeating state. 
		\item Those ECA that remain in a random state. 
		\item Those ECA that combine elements of randomness and predictability ("complex"). 
	\end{enumerate}
	
	While these are qualitative designations as opposed to quantitative ones, how emergence is distributed over representatives of these four classes will go a long way to providing intuitive insight into what kind of behaviours support emergence and which do not.
	
	The ECA can also be explored at a variety of scales, corresponding to the number of cells. For each rule, we created eight state-transition graphs, each one corresponding to a size of five cells (with a 32-vertex state-transition graph) to 12 cells (corresponding to a 4096-vertex state-transition graph). This provides a unique opportunity to explore how the size of a system contributes to it's determinism, degeneracy, and capacity to support emergence while holding the generating dynamics constant. While there are 256 possible rules, quite a few of them are identical: there are in fact only 88 "unique" rules (all others being symmetrical to at least one other rule), so for these results here, we picked only those 88 distinct rules (the numbers for each rule can be found in Supplementary Datasets). 
	
	\subsection{Rossler Attractor}
	\label{sec:rossler}
	The Rossler attractor is a three-dimensional dynamical system commonly used as a toy-model when exploring chaotic dynamics. It is defined by:
	
	\[{\frac{dx}{dt}} = -y - z\]
	\[{\frac{dy}{dt}} = x + ay\]
	\[{\frac{dz}{dt}} = b + z(x - c)\]
	
	where \textit{a, b} and \textit{c} are free parameters that control the dynamics of the system. For this study, we held the values of $b$ and $c$ constant at 2 and 4 respectively and varied the value of $a$ within a range of 0.37 - 0.43 in increments of 0.001, following Myers et al., (2019). From the attractor we took only the $x$-series, giving us a large set of time-series corresponding to dynamics in period, critical, and chaotic regimes. 
	\subsubsection{Ordinal Partition Network}
	\label{sec:opns}
	The ordinal partition network (OPN) represents a discretized approximation of a continuous attractor. Based on the notion of permutation embedding, each node represents a pattern of activity and the weights between the nodes represent the probabilities that, given some current state $S_i$, at the next timestep, the system evolves into $S_j$. Constructing the OPN is reasonably simple and requires choosing only two free parameters: the embedding dimension $d$, and the temporal lag $\tau$. Given some real-valued time-series $X = [x_1, x_2, x_3, x_4, x_5 ... x_n]$, we construct a series of $d$-dimensional vectors: $V = [x_\sigma, x_{\tau\sigma}, x_{2\tau\sigma} ... x_{d\tau\sigma}]. $ We then find the ordinal permutation of $V$, which is the rank of each element in the vector. For example, suppose $V = [0.1, -0.1, 0.5, 0.2, -0.9]$. The ordinal permutation of $\pi(V)$ would be $[2,1,4,3,0]$, which is the indices of $V$ that would sort it. 
	
	The primary benefit of the permutation embedding is that it allows us to map every possible real-valued vector $V$ into one of only $d!$ possible ordinal partitions. To construct the state-transition network we count how many times $\pi(V_i)$ is followed by $\pi(V_j)$ and calculate the $P(\pi(V_j) | \pi(V_i))$. The result is a Markovian network with a finite number of nodes, and for which the out-degrees of each node define a probability distribution. A random walk on this network returns a plausible series of ordinal permutations which could conceivably be reconstructed into a probable continuous time-series. 
	
	The question of what are the optimal values of $d$ and $\tau$ are contentious in the literature. In the absence of a "best practice" for the Rossler attractor we used the same embedding as Myers et al (2019): $d=6, \tau=40$.
	
	\subsection{Software}
	
	All the relevant code can be found in the associated GitHub repository: \newline https://github.com/thosvarley/causal\_emergence
	
	Hosted therein is the source code for the ECA, the Rossler system, and the causal emergence package (written in Cython) as well as the relevant analysis scripts. The actual data (the state-transition graphs for all the ECA and the Rossler system) are not included due to hosting limits. All analyses were done in Python 3.7 and Cython \cite{behnel_cython:_2011}. Other packages used include the Numpy library (version 1.15.4) \cite{walt_numpy_2011}, Scipy (version 1.3.1) \cite{jones_scipy_2001}, Scikit-Learn (version 0.20.0) \cite{pedregosa_scikit-learn:_2011}, Matplotlib (version 2.2.2), \cite{hunter_matplotlib:_2007}, Spyder (version 3.2.3), iGraph (version 0.7.1) \cite{csardi_igraph}. Analysis was done in the Anaconda Python Environment (version 5.0.0). 
	
	\section{Results \& Discussion}
	\subsection{Elementary Cellular Automata}
	
	\begin{figure}
		\centering
		\includegraphics[scale=0.65]{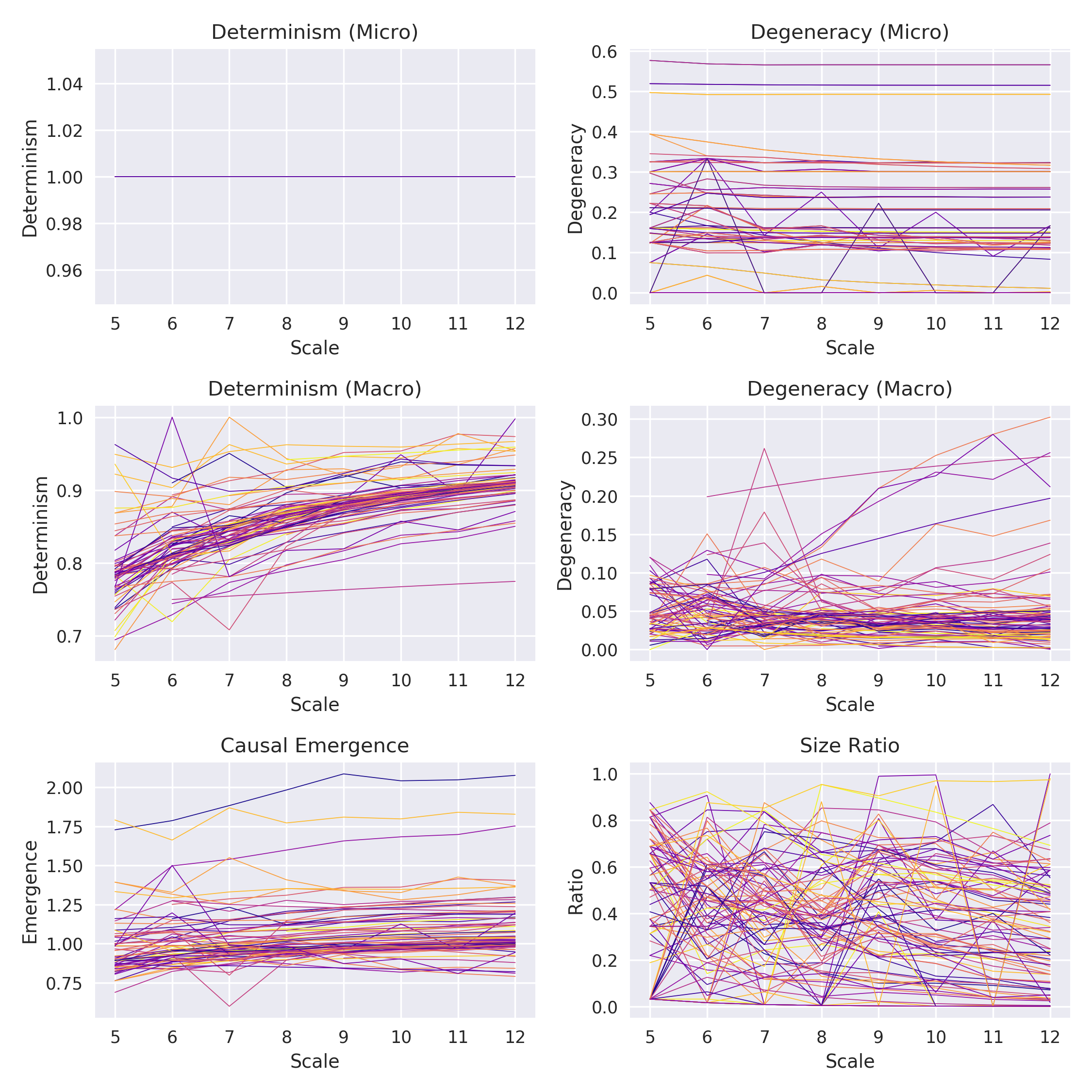}
		\caption{How the components of causal emergence (determinism, degeneracy, and effectiveness, at micro and macro-scales) relate to the size of the state-transition graph for each of the 88 unique ECA at all 7 scales. Note that as the systems get larger, macro-networks become more deterministic, while degeneracy seems to be largely unaffected. There may be a subtle trend that systems with a larger number of state are capable of displaying slightly higher emergence as well.}
		\label{fig:scales}
	\end{figure}

	\begin{figure}
		\centering
		\includegraphics[scale=0.65]{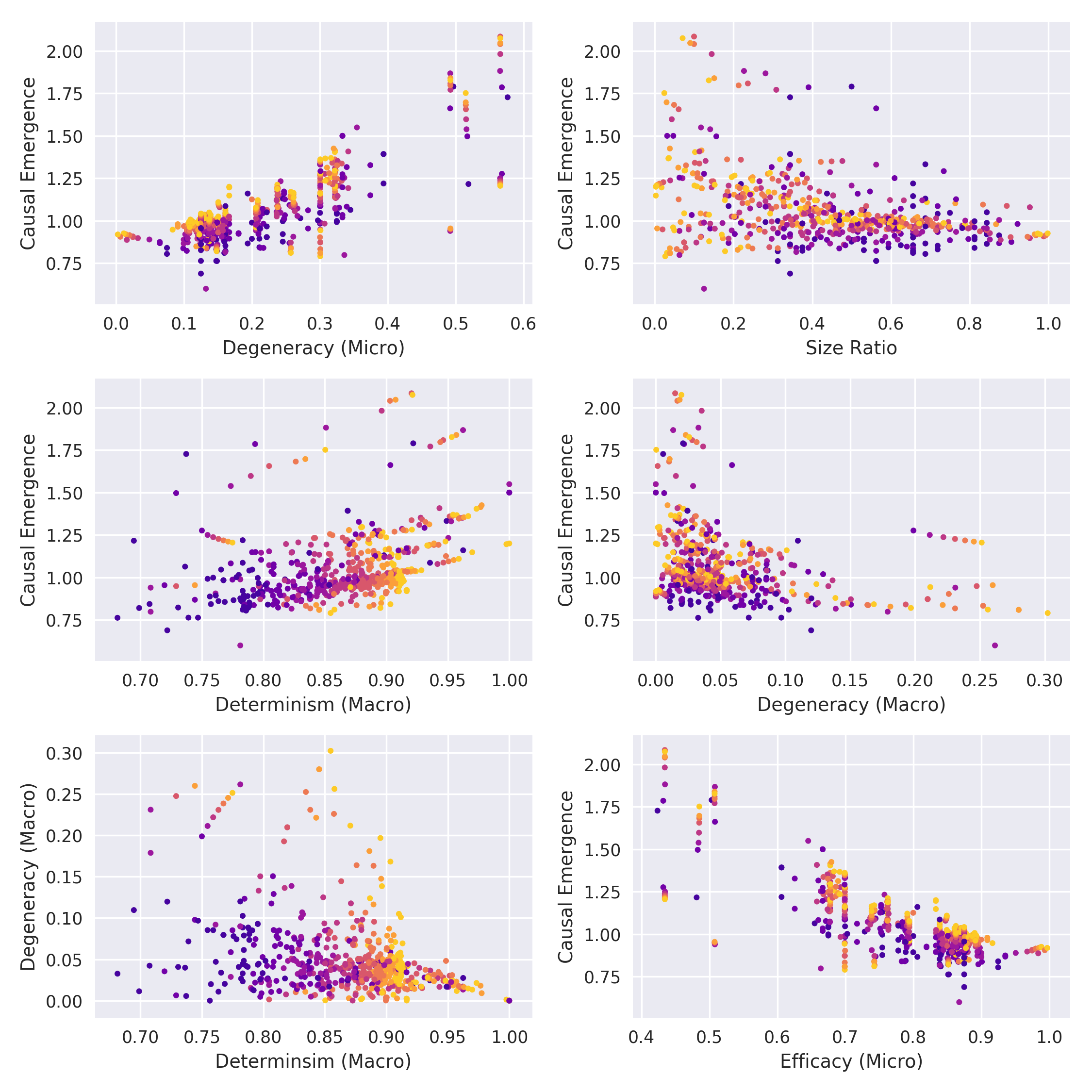}
		\caption{How the components of causal emergence (determinism, degeneracy, and effectiveness, at micro and macro-scales) relate to each-other. Note that as micro-scale degeneracy grows, capacity for causal-emergence climbs, while as the macro-degeneracy grows, capacity for causal emergence falls. Macro-scale determinism and degeneracy do not appear to be completely independent either.}
		\label{fig:corr}
	\end{figure}

	\begin{figure}
		\centering
		\includegraphics[scale=0.4]{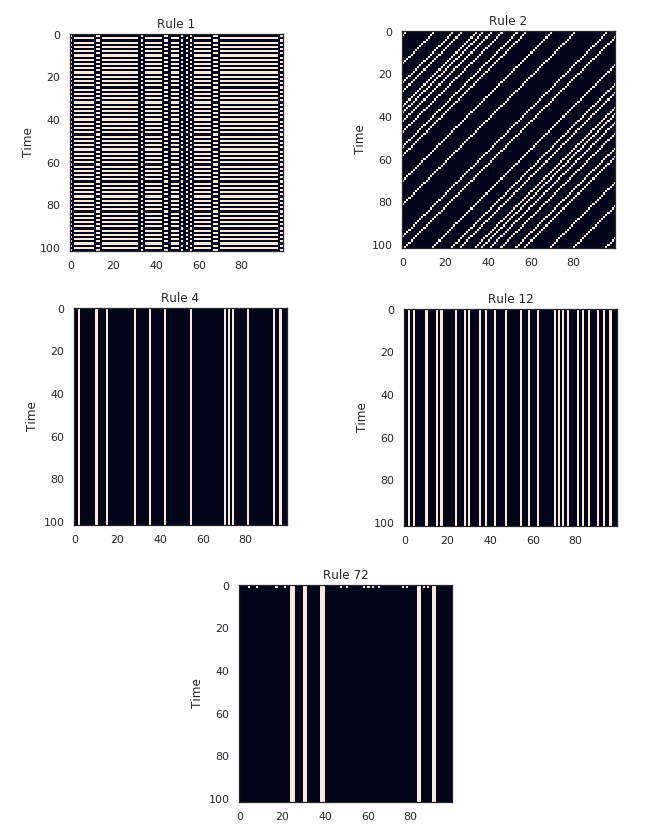}
		\caption{The five rules that displayed the highest causal emergence over the whole range of scales, using the same random initial condition for each rule. None of these rules are particularly exciting, in terms of long-term behaviour.}
		\label{fig:eca}
	\end{figure}

	\begin{figure}
		\centering
		\includegraphics[scale=0.1]{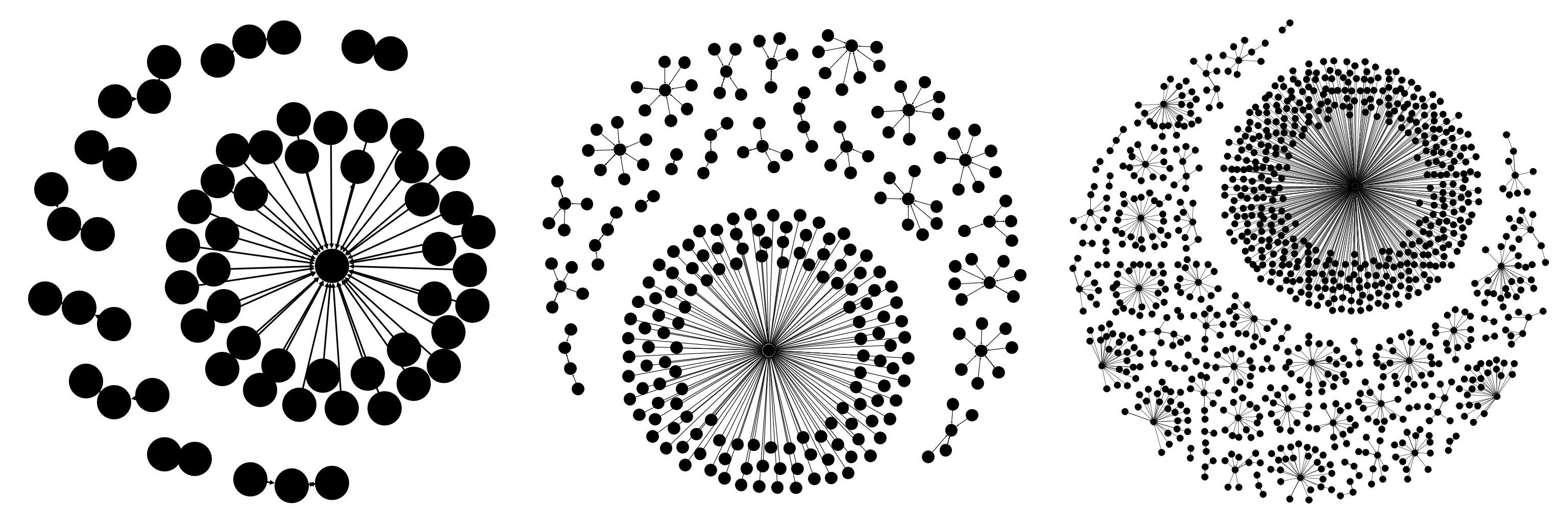}
		\includegraphics[scale=0.1]{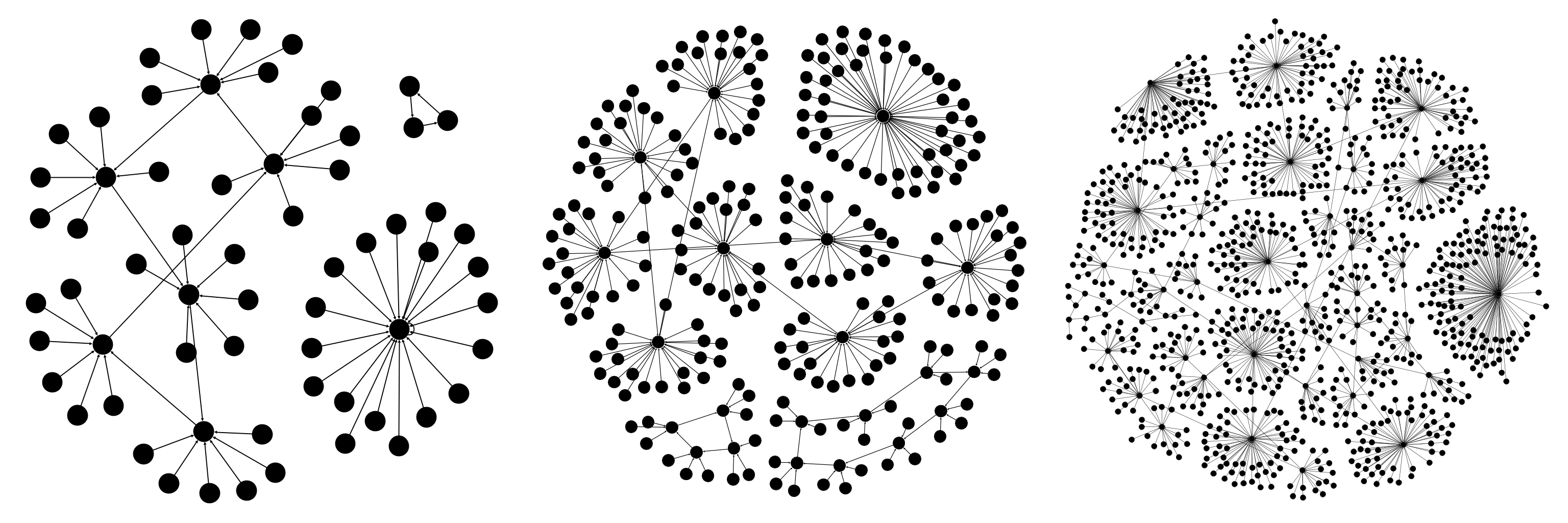}
		\includegraphics[scale=0.1]{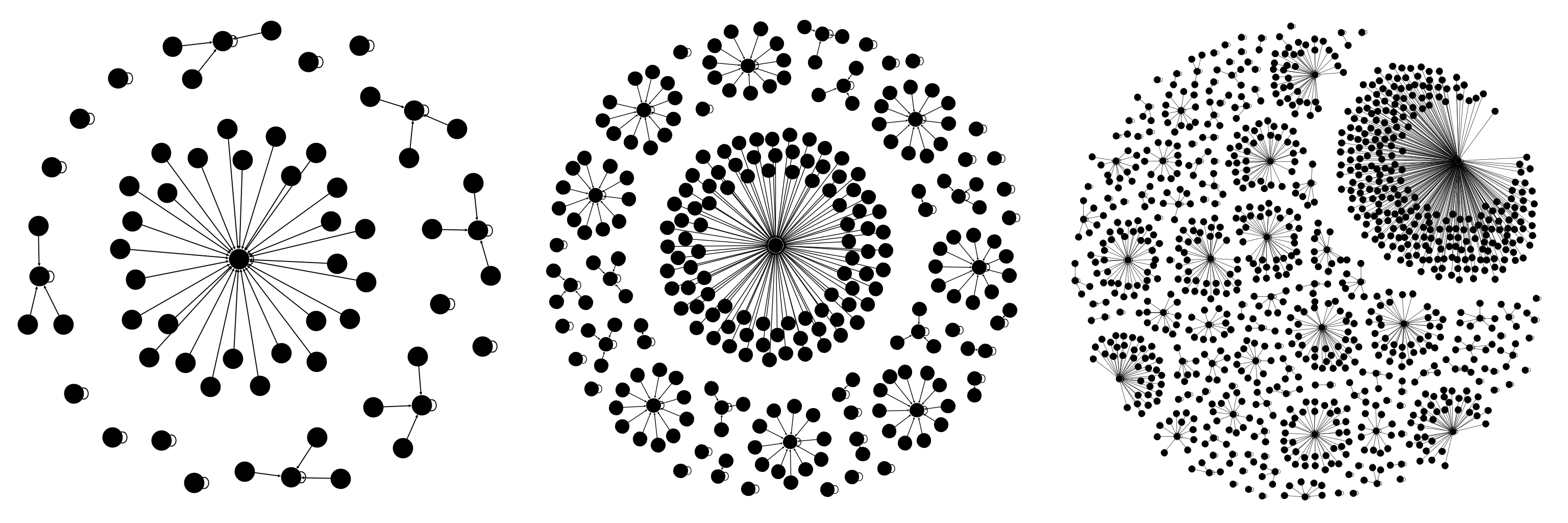}
		\caption{The state-transition graphs for the three rules that showed the greatest causal emergence.  Rule 1 (top row), rule 2 (middle row), and rule 4 (bottom row), when the system is comprised of 6 elements (left column), 8 elements (middle column) and 10 elements (right column). When visualizing the state transition graphs, it is clear that all of them contain a large number of star-like motifs that are easily collapsed under renormalization. }
		\label{fig:stgs}
	\end{figure}
	
	Approximately 8\% of the rules had state-transition graphs that collapsed down to a single point (where causal emergence is undefined) across all scales. These rules were:
	
	\[\{8, 15, 45, 51, 170, 204\}\]
	
	This set of rules shows a range of behaviours: rule 45 is one of the canonically "interesting" rules, showing complex, chaotic behaviour, while rule 8 evolves to a uniform steady state, and the rest fall into oscillatory modes. This suggests that this kind of collapse is not restricted to a single long-term behaviour. 
	
	Approximately 43\% of the rules displayed true causal emergence on average across the range of all scales, while approximately 49\% showed causal reduction on average. As with the rules that collapsed, there was not an obvious relationship between rule class and causal emergence. Of the 17 rules typically described as being class 3 or class 4 (showing complex, non-oscillatory behaviour), approximately 30\% showed true causal emergence when averaged across scales, while 70\% showed causal reduction (for a list of the 17 "interesting" rules, see Supplementary Datasets). An interesting example of these differences is to compare rules 30, 60, and 90, all of which are commonly cited as ECA exhibiting nontrivial behaviour. Rule 90 had a mean causal emergence across scales of 1.05, while rules 30 and 60 showed causal reduction, with mean values of 0.89 and 0.94 respectively. Rule 110, another nontrivial ECA, with behaviour described as on the critical boundary between periodicity and chaos also showed causal reduction, with a mean value of 0.94 across scales. 
	
	Interestingly, the rules that showed the highest causal emergence generally had fairly trivial long-term behaviours. The top five rules with the highest average causal emergence were (in descending order):
	
	\[\{1,4,2,72,12\}\]
	
	With mean values ranging from 1.95 for rule 1 down to 1.34 for rule 12 (for visualizations of the space-time diagrams for these fives rules, see Figure \ref{fig:eca}). By examining the state-transition graphs of these rules, we can see that high CE seems to be associated with a large number of communities arranged in a star, or hub-and-spoke motifs. This is consistent with the intuition: all spokes on a star motif lead to the same attractor state and as a result, are causally equivalent and can be collapsed without loss of information about the future of the system. In contrast, the systems that displayed the least CE (but did not collapse to single points) had tree-like state transition graphs (data not shown).
	
	In general, causal emergence was largely consistent across scales, which we take as a promising sign that the algorithm proposed by Griebenow et al., \cite{griebenow_finding_2019} is robust: the same system at different sizes returns largely the same results. Unexpectedly, however, despite the relative constancy of emergence, the macro-scale determinism grew as the size of the network increased, while degeneracy at the macro scale was generally constant or trending towards zero as the network size increased (with a few notable exceptions). 
	
	In the ECA, we were also able to see how the components of causal emergence (micro-scale determinism and degeneracy, as well as macro-scale components) were related to each-other (see Fig. \ref{fig:scales}). Unsurprisingly, as micro-scale degeneracy increased, the capacity of the system to support CE increased. This is consistent with the notion that degeneracy at the micro-scale allows for efficient coarse-graining and the emergence of informationally rich higher scales. There was also a positive relationship between the macro-scale determinism and CE, which was mirrored by a negative relationship between macro-scale degeneracy and CE. This too is consistent with the notion that CE is an increase in effectiveness at the macro-scale relative to the micro-scale: as effectiveness is highest when determinism is high and degeneracy is low, an increase in macro-scale determinism and a decrease in macro-scale degeneracy corresponds to greater CE. This is also reflected in a strong negative correlation between micro-scale effectiveness and causal emergence. 
	
	These results provide an empirical verification of the theory of causal emergence and detail how different dynamical systems (corresponding to the 88 unique elementary cellular automata), with different long-term behaviour, display unique combinations of determinism, degeneracy, and effectiveness at macro- and micro-scales. 
	
	\subsection{Rossler Attractor}
	
\begin{figure}
		\centering
		\includegraphics[scale=0.125]{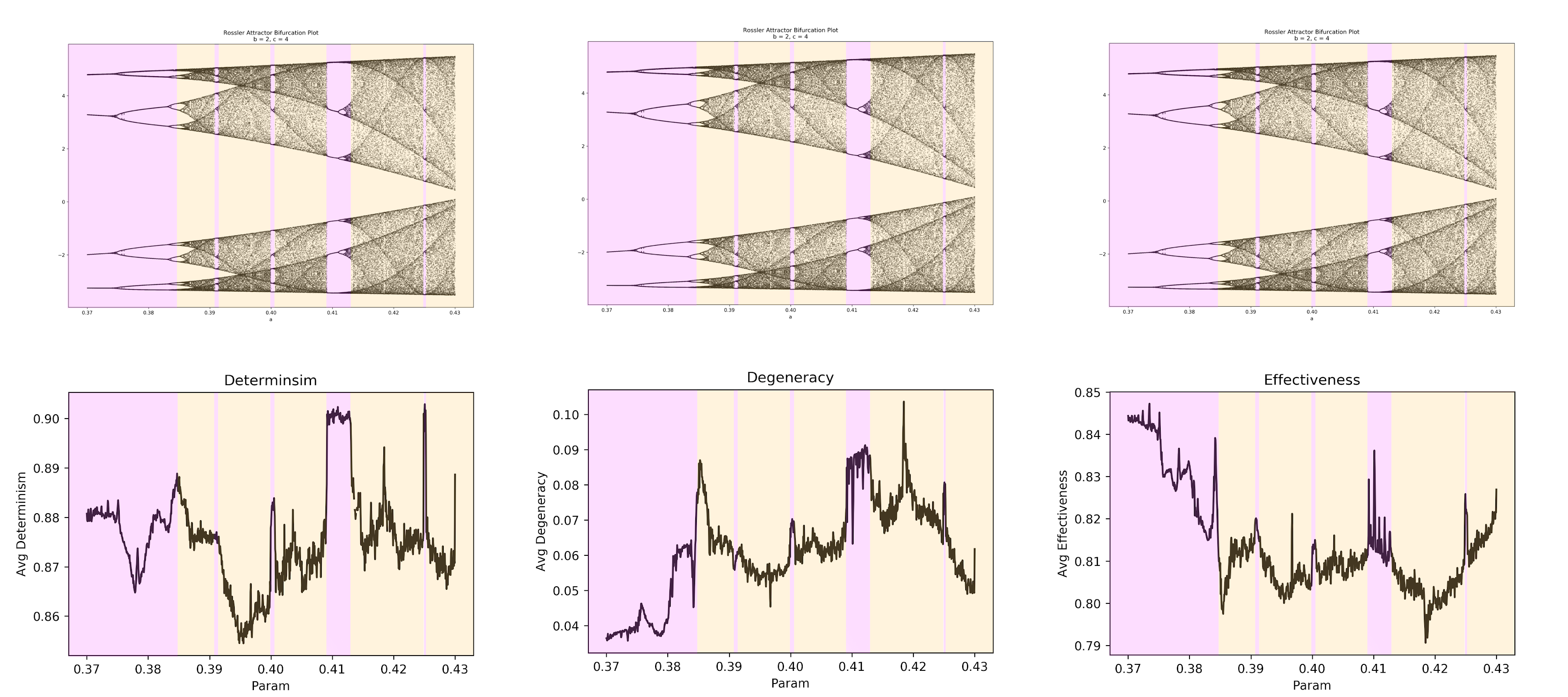}
		\caption{How the determinism (left), degeneracy (middle), and efficacy (right) change as the Rossler attractor is swept across the period doubling cascade and onset of deterministic chaos. In the Rossler equations, $b$=2, $c$=4, and $a$ is the independent variable. It is clear that all three components of causal emergence are sensitive to changes in the dynamical regime of the system. 
			 }
		\label{fig:det_deg}
	\end{figure}

	The ordinal partition network (OPN) of the Rossler System time series provides a natural way to bring information theoretic analysis to bear on on a continuous dynamical system, as detailed in \cite{mccullough_time_2015,myers_persistent_2019}. By fixing $b=2,c=4$ and sweeping through $a$ from a range of 0.37-0.43, we can observe how CE changes as the system undergoes a bifurcation cascade leading up to a critical phase transition. We found that determinism underwent a brief climb, followed by a significant drop immediately following the first bifurcation and then generally climbed until the onset of the period-doubling cascade, at which point, it collapsed precipitously. In general, chaotic dynamics were associated with lower levels of determinism. Upon brief transitions back into periodic dynamics, the determinism climbed dramatically, resulting in "deterministic plateaus" surrounded by low-determinism, chaotic dynamics. For visualization, see Figure \ref{fig:det_deg}.
	
	Degeneracy displayed a similar pattern to determinism. At the onset of the period doubling cascade, the degeneracy spiked before falling. As with determinism, the degeneracy seemed to increase any time the system transitioned into a periodic regime, collapsing following the transition back to chaos. Interestingly, the degeneracy" appears to "anticipate" (at risk of anthropomorphizing a mathematical construct) the transition back to determinism and begins to rise prior to the significant leap that occurs on the onset of determinism. At this point, we do not know exactly what dynamical process it producing this effect. 
	
\begin{figure}
		\centering
		\includegraphics[scale=0.15]{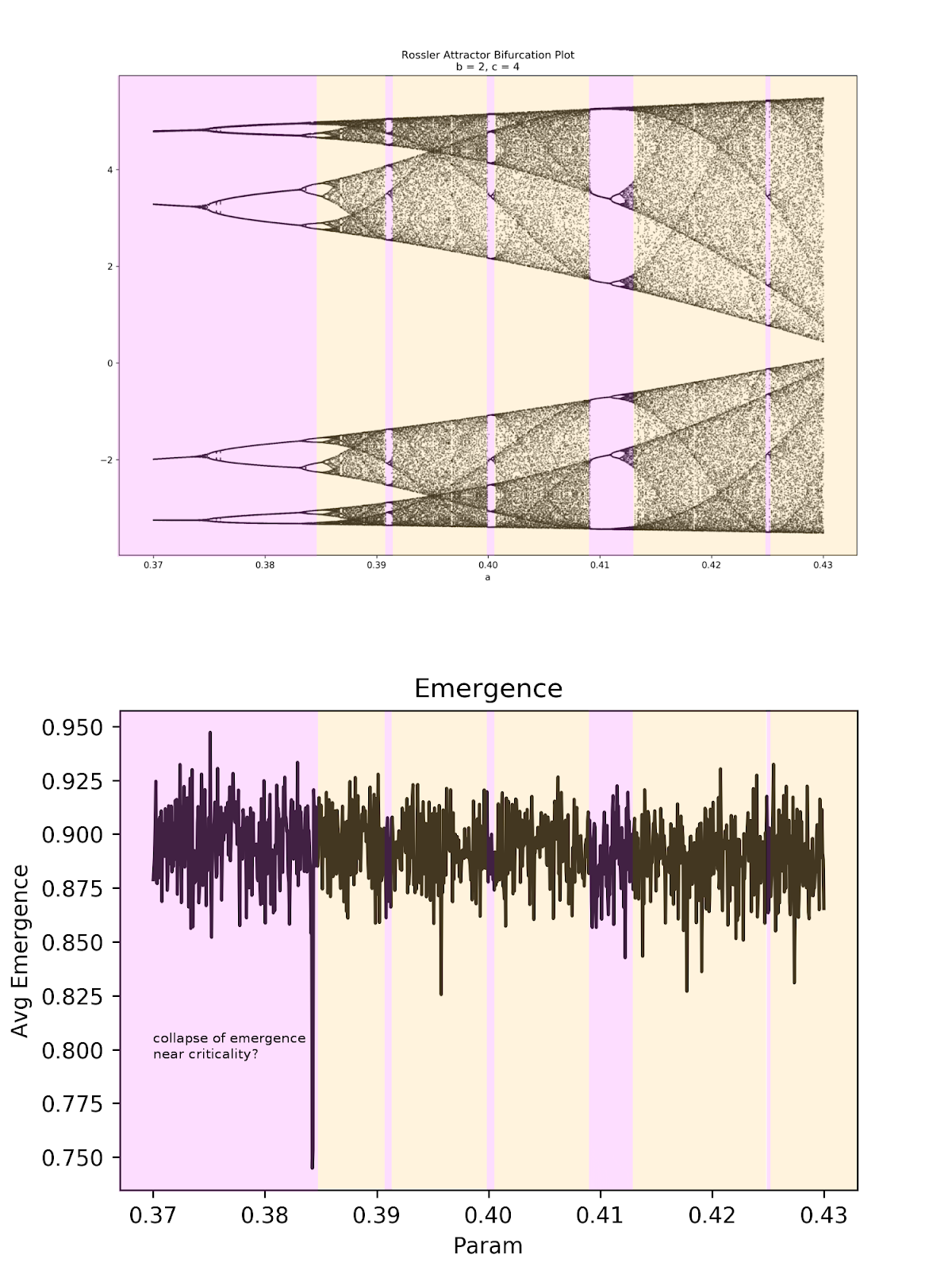}
		\includegraphics[scale=0.425]{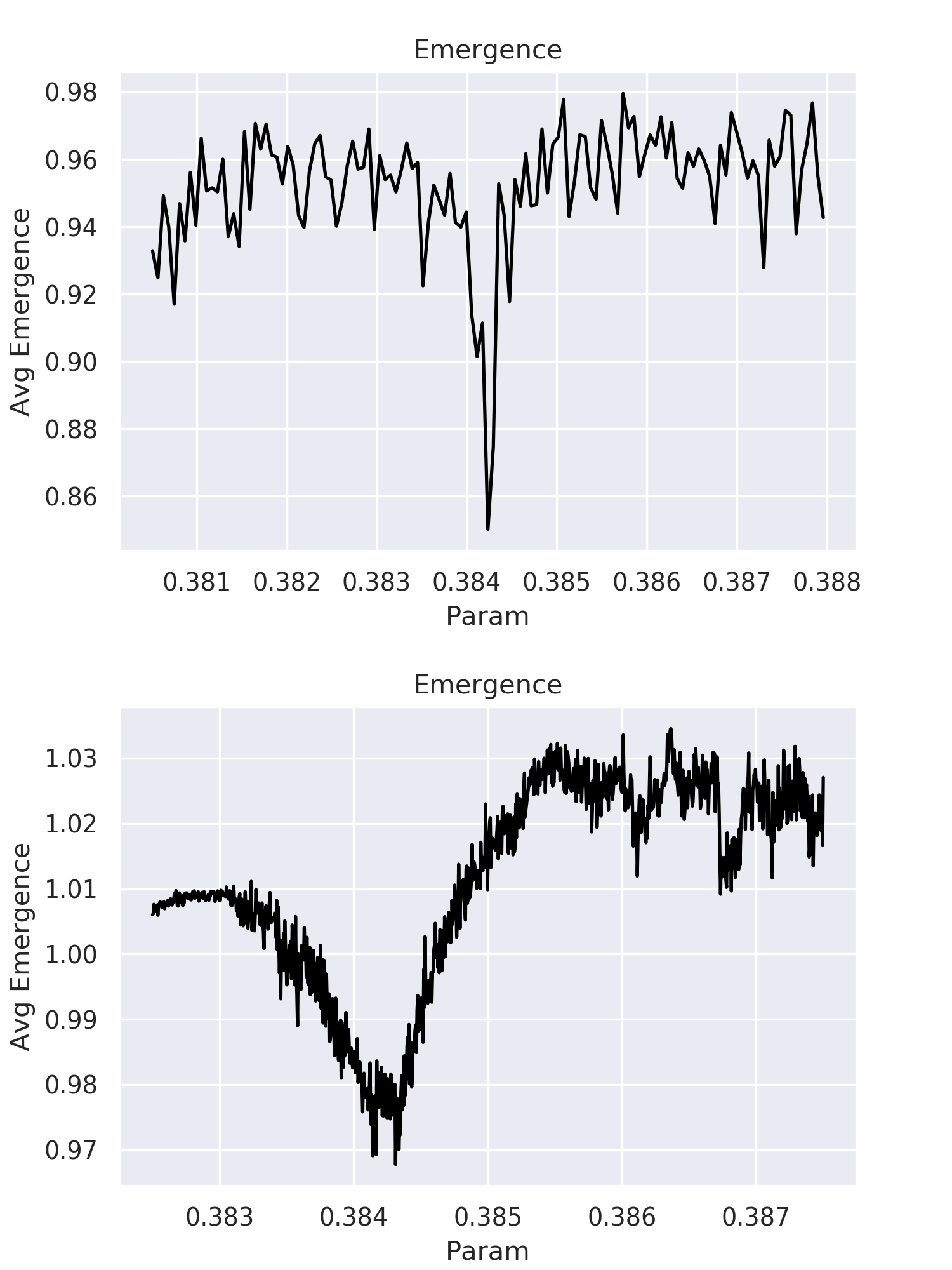}
		\caption{The change in causal emergence calculated from ordinal partition networks as the Rossler system is swept through the onset of chaos. Left: Notice the significant decrease in emergence that occurs near exactly on the onset of the period doubling cascade. Right, top: this is the same as the bottom left figure, but zoomed in on the period doubling cascade. Notice that the drop includes 4-5 distinct points, suggesting that it is not an artefact. Right, bottom: to test whether the drop was robust, we re-ran the analysis, with a higher sampling rate covering just the period doubling cascade. The drop, reaching it's minimum at ~0.384 persists, although it is not nearly as significant in terms of absolute magnitude.}
		\label{fig:eff_Ce}
	\end{figure}

	The effectiveness, being the difference between determinism and degeneracy, displayed largely similar patterns as determinism: a spike upon the onset of the period doubling cascade, followed by a dramatic decrease. As both determinism and degeneracy spiked during periodic regimes, the effectiveness was necessarily lower, although the transitions from chaos to periodicity were still marked by increases in value. For visualization see \ref{fig:eff_Ce}. Causal emergence was arguably the least interesting of the measures reported here. It remained noisy around a relatively constant value, with the notable exception that it plunged upon the onset of the period doubling cascade. We might have have expected that emergence may climbed at the onset of the phase transition, as many have proposed that complex, emergent phenomena emerge at the "Edge of Chaos." 
	
	\section{Conclusion}
	In this paper, we present how the causal emergence framework, first introduced by Hoel et al., \cite{hoel_quantifying_2013,hoel_can_2016}, performs when analysing toy-models of discrete and continuous dynamical systems. The discrete system we used were the 88 unique elementary cellular automata, and we found that the set of rules displayed a wide distribution of determinism, degeneracy, and emergence. Interestingly, the rules that displayed the consistently highest values of emergence were not the ones that displayed the most visually interesting long-term dynamics. To explore emergence in continuous dynamical systems, we constructed discretized state-transition networks for a Rossler system that we swept through a period doubling cascade. We found that micro-scale determinism, degeneracy, and efficacy changed significantly depending on the dynamical regime and seemed to be sensitive to changes in dynamics prior to critical phase transitions. Upon the onset of the period doubling cascade, the capacity for the system to support higher-level causal emergence decreased dramatically. 
	
	\section*{Acknowledgements}
	I would like to thank Dr. YY Ahn, Dr. Randall Beer, Dr. Olaf Sporns, and Dr. Alice Patania for their thoughtful insights and discussions over the course of these projects. I would also like to thank Ross Griebenow for assistance coding the spectral clustering algorithm, and Dr. Eric Hoel for helping me understand the intuition behind causal emergence. 
	
	\bibliographystyle{apalike}
	\bibliography{emergence.bib}
\end{document}